\documentclass{llncs}

\usepackage[usenames]{color}

\usepackage{latexsym}
\usepackage{amsfonts}
\usepackage{graphicx}
\usepackage{amsmath}
\usepackage{amssymb}
\begin{document}

\title{Public key exchange using semidirect product\\
 of (semi)groups}
\author{Maggie Habeeb\inst{1} \and Delaram Kahrobaei\inst{2}  \and Charalambos Koupparis\inst{3} \and Vladimir Shpilrain\inst{4}}
\institute{California University of Pennsylvania\\
\email{habeeb@calu.edu} \thanks{Research of Maggie Habeeb was
partially supported  by the NSF-LSAMP fellowship.} \and  CUNY
Graduate Center and City Tech, City University of New York
\email{dkahrobaei@gc.cuny.edu} \thanks{Research of Delaram Kahrobaei
was partially supported by a PSC-CUNY grant from the CUNY research
foundation, as well as the City Tech foundation. Research of Delaram
Kahrobaei and Vladimir Shpilrain was also supported by the ONR
(Office of Naval Research) grant N000141210758.} \and CUNY
Graduate Center, City University of New York\\
\email{ckoupparis@gc.cuny.edu} \and The City College of New York and
CUNY Graduate Center \email{shpil@groups.sci.ccny.cuny.edu}
\thanks{Research of Vladimir Shpilrain was partially supported by the NSF grants DMS-0914778  and CNS-1117675.}}

%

\maketitle

\begin{abstract}
In this paper, we   describe a brand new  key exchange protocol
based on a semidirect product of (semi)groups (more specifically, on
extension of a (semi)group by automorphisms), and then focus on
practical instances of this general idea. Our protocol can be based
on any group, in particular on any non-commutative group. One of its
special cases is the standard Diffie-Hellman protocol, which is
based on a cyclic group. However, when our protocol is used with a
non-commutative (semi)group, it acquires several useful features
that make it compare favorably to the Diffie-Hellman protocol. Here
we also suggest a particular non-commutative semigroup (of matrices)
as the platform and show that security of the relevant protocol is
based on a quite different assumption compared to that of the
standard Diffie-Hellman protocol.

\end{abstract}

\section{Introduction}

 It is rare that the beginning of a whole new area of science can
be traced back to one particular paper. This is the case with public
key cryptography; it started  with the seminal paper \cite{DH}.

The simplest, and original, implementation of the protocol uses the
multiplicative group of integers modulo $p$, where $p$ is prime and
$g$ is primitive $\mod p$. A more general description of the
protocol uses an arbitrary finite cyclic group.

\begin{enumerate}

\item Alice and Bob agree on a finite cyclic group $G$ and a generating element $g$ in $G$.
 We will write the group $G$ multiplicatively.

\item Alice picks a random natural number $a$ and sends $g^a$ to Bob.

\item    Bob picks a random natural number $b$ and sends $g^b$ to Alice.

\item   Alice computes $K_A=(g^b)^a=g^{ba}$.

\item  Bob computes $K_B=(g^a)^b=g^{ab}$.
\end{enumerate}

Since $ab=ba$, both Alice and Bob are now in possession of the same
group element $K=K_A= K_B$ which can serve as the shared secret key.

The protocol is considered secure against eavesdroppers if $G$ and
$g$ are chosen properly. The eavesdropper must solve the {\it
Diffie-Hellman problem} (recover $g^{ab}$ from $g$, $g^a$ and $g^b$)
to obtain the shared secret key. This is currently considered
difficult for a ``good" choice of parameters (see e.g.
\cite{Menezes} for details).

There is  an ongoing search for other platforms where the
Diffie-Hellman or similar key exchange could be carried out more
efficiently, in particular with public/private keys of   smaller
size. This search already gave rise to several interesting
directions, including a whole area of elliptic curve cryptography.
We also refer the reader to \cite{MSU} for a survey of  proposed
cryptographic primitives based on non-abelian (= non-commutative)
groups. A survey of these efforts is outside of the scope of the
present paper; our goal here is to suggest a new  key exchange
protocol based  on extension of a (semi)group by automorphisms. Our
protocol can be based on any group, in particular on any
non-commutative group. It has some superficial resemblance to the
classical Diffie-Hellman protocol, but there are several distinctive
features that, we believe, give our protocol important advantages.
In particular, even though the parties do compute a large power of a
public element (as in the classical Diffie-Hellman protocol), they
do not transmit the whole result, but rather just part of it.

We also  describe in this paper some particular instances of our
general protocol.  In particular, we  suggest a non-commutative
semigroup (of matrices) as the platform and show that security of
the relevant protocol is based on a quite different assumption
compared to that of the standard Diffie-Hellman protocol.

We mention another, rather different, proposal \cite{PHKCP} of a
cryptosystem based on the semidirect product of two groups and yet
another, more complex, proposal of a key agreement based on the
semidirect product of two monoids \cite{AAGL}. Both these proposals
are very different from ours. Also, the extended abstract
\cite{HKS}, despite the similarity of the title, has very little
overlap with the present paper. In particular, the key exchange
protocol in Section \ref{protocol_holomorph} of the present paper is
brand new.

Finally, we note that the basic construction (semidirect product) we
use in this paper can be adopted, with some simple modifications, in
other algebraic systems, e.g. associative rings or Lie rings, and
key exchange protocols similar to ours can be built on those.

\section{Semidirect products and extensions  by automorphisms}
\label{Semidirect}

We include this section to make the exposition more comprehensive.
The reader who is uncomfortable with group-theoretic constructions
can skip to subsection \ref{holomorph}.

We now recall the definition of a semidirect product:

\begin{definition} Let $G, H$ be two groups, let $Aut(G)$ be the group of automorphisms of $G$,
and let $\rho: H \rightarrow Aut(G)$ be a homomorphism. Then the
semidirect product of $G$ and $H$ is the set
$$\Gamma = G \rtimes_{\rho} H = \left \{ (g, h): g \in G, ~h \in H \right \}$$
with the group operation given by\\
\centerline{$(g, h)(g', h')=(g^{\rho(h')} \cdot  g', ~h \cdot h')$.}\\
Here $g^{\rho(h')}$ denotes the image of  $g$ under the automorphism
$\rho(h')$, and when we write a product $h \cdot h'$ of two
morphisms, this means that $h$ is applied first.
\end{definition}

In this paper, we focus on a special case of this construction,
where the group $H$ is just a subgroup of the group $Aut(G)$. If
$H=Aut(G)$, then the corresponding semidirect product is called the
{\it holomorph} of the group $G$. We give some more details about
the holomorph in our Section \ref{holomorph}, and in Section
\ref{protocol_holomorph} we describe a key exchange protocol that
uses (as the platform) an extension of a group $G$ by a {\it cyclic}
group of automorphisms.

\subsection{Extensions  by automorphisms} \label{holomorph}

A particularly simple special case of the semidirect product
construction is where the group $H$ is just a subgroup of the group
$Aut(G)$. If $H=Aut(G)$, then the corresponding semidirect product
is called the {\it holomorph} of the group $G$. Thus,  the holomorph
of $G$, usually denoted by $Hol(G)$, is the set of all pairs $(g,
~\phi)$, where $g \in G, ~\phi \in Aut(G)$, with the group operation
given by ~$(g,~\phi)\cdot  (g',~\phi') = (\phi'(g)\cdot g',~\phi
\cdot \phi')$.

It is often more practical to use a subgroup of $Aut(G)$ in this
construction, and this is exactly what we do in Section
\ref{protocol_holomorph}, where we describe a key exchange protocol
that uses (as the platform) an extension of a group $G$ by a cyclic
group of automorphisms.

\begin{remark}
One can also use this construction if $G$ is not necessarily a
group, but just a semigroup, and/or consider endomorphisms of $G$,
not necessarily automorphisms. Then the result will be a semigroup;
this is what we use in our Section \ref{Matrices}.
\end{remark}

\section{Key exchange protocol}
\label{protocol_holomorph}

In the simplest implementation of the construction described in our
Section \ref{holomorph}, one can use just a cyclic subgroup (or a
cyclic subsemigroup) of the group $Aut(G)$ (respectively, of the
semigroup  $End(G)$ of endomorphisms) instead of the whole group of
automorphisms of $G$.

Thus, let $G$ be a (semi)group. An element $g\in G$ is chosen and
made public as well as an arbitrary automorphism $\phi\in Aut(G)$
(or an arbitrary endomorphism $\phi\in End(G)$). Bob chooses a
private $n\in \mathbb{N}$, while Alice chooses a private $m\in
\mathbb{N}$. Both Alice and Bob are going to work with elements of
the form $(g, \phi^r)$, where $g\in G, ~r\in \mathbb{N}$. Note that
two elements of this form are multiplied as follows:  ~$(g, \phi^r)
\cdot (h, \phi^s) = (\phi^s(g)\cdot h, ~\phi^{r+s})$.
\medskip

\begin{enumerate}

\item Alice computes $(g, \phi)^m = (\phi^{m-1}(g) \cdots \phi^{2}(g) \cdot \phi(g) \cdot g,
~\phi^m)$ and sends {\bf only the first component} of this pair to
Bob. Thus, she sends to Bob {\bf only} the element $a =
\phi^{m-1}(g) \cdots \phi^{2}(g) \cdot \phi(g) \cdot g$ of the
(semi)group $G$.
\medskip

\item Bob computes $(g, \phi)^n = (\phi^{n-1}(g) \cdots \phi^{2}(g) \cdot \phi(g) \cdot g,
~\phi^n)$ and sends {\bf only the first component} of this pair to
Alice. Thus, he sends to Alice {\bf only} the element $b =
\phi^{n-1}(g) \cdots \phi^{2}(g) \cdot \phi(g) \cdot g$ of the
(semi)group $G$.
\medskip

\item Alice computes $(b, x) \cdot (a, ~\phi^m) = (\phi^m(b) \cdot a,
~x  \cdot \phi^{m})$. Her key is now $K_A = \phi^m(b) \cdot a$. Note
that she does not actually ``compute" $x \cdot \phi^{m}$ because she
does not know the automorphism $x=\phi^{n}$; recall that it was not
transmitted to her. But she does not need it to compute $K_A$.
\medskip

\item Bob computes $(a, y) \cdot (b, ~\phi^n) = (\phi^n(a) \cdot b,
~y \cdot \phi^{n})$. His key is now $K_B = \phi^n(a) \cdot b$.
Again, Bob does not actually ``compute" $y \cdot \phi^{n}$ because
he does not know the automorphism $y=\phi^{m}$.
\medskip

\item Since $(b, x) \cdot (a, ~\phi^m) = (a, ~y) \cdot (b, ~\phi^n) =
(g, ~\phi)^{m+n}$, we should have $K_A = K_B = K$, the shared secret
key.

\end{enumerate}

\begin{remark}
Note that, in contrast with the ``standard" Diffie-Hellman key
exchange, correctness here is based on the equality $h^{m}\cdot
h^{n} = h^{n} \cdot h^{m} =  h^{m+n}$  rather  than on the equality
$(h^{m})^{n} = (h^{n})^{m} = h^{mn}$. In  the ``standard"
Diffie-Hellman set up, our trick would not work because, if the
shared key $K$ was just the product of two  openly transmitted
elements, then anybody, including the eavesdropper, could compute
$K$.
\end{remark}

\section{Computational cost}
\label{cost}

From the look of transmitted elements in our protocol in Section
\ref{protocol_holomorph}, it may seem that the parties have to
compute a product of $m$ (respectively, $n$) elements of the
(semi)group $G$. However, since the parties actually compute powers
of an element of $G$, they can use the ``square-and-multiply"
method, as in the standard Diffie-Hellman protocol. Then there is a
cost of applying an automorphism $\phi$ to an element of $G$, and
also of computing powers of $\phi$. These costs depend, of course,
on a specific platform (semi)group that is used with our protocol.
In our first, ``toy" example (Section \ref{Toy} below), both
applying an automorphism $\phi$ and computing its powers amount to
exponentiation of elements of $G$, which can be done again by the
``square-and-multiply" method. In our main example, in Section
\ref{Matrices}, $\phi$ is a conjugation, so applying $\phi$ amounts
to just two multiplications of elements in $G$, while computing
powers of $\phi$ amounts to exponentiation of two elements of $G$
(namely, of the conjugating element and of its inverse).

Thus, in either instantiation of our protocol considered in this
paper, the cost of computing $(g, \phi)^n$ is $O(\log n)$, just as
in the standard Diffie-Hellman protocol.

%

\section{``Toy example": multiplicative $\mathbb{Z}_p^*$}
\label{Toy}

As one of the simplest instantiations  of our protocol, we use here
the multiplicative group $\mathbb{Z}_p^*$  as the platform group $G$
to illustrate what is going on. In selecting a prime $p$, as well as
private exponents $m, n$, one can follow the same guidelines as in
the ``standard" Diffie-Hellman.

Selecting the (public) endomorphism  $\phi$ of the group
$\mathbb{Z}_p^*$ amounts to selecting yet another integer $k$, so
that for every $h \in \mathbb{Z}_p^*$, one has $\phi(h) = h^k$. If
$k$ is  relatively prime to $p-1$, then $\phi$ is actually an
automorphism. Below we assume that $k > 1$.

Then, for an element $g \in \mathbb{Z}_p^*$, we have:
$$(g, \phi)^m =
(\phi^{m-1}(g) \cdots    \phi(g) \cdot \phi^{2}(g) \cdot g,
~\phi^m).$$

We focus on the first component of the element on the right; easy
computation shows that it is equal to $g^{k^{m-1} +\ldots +k +1} =
g^{\frac{k^{m}-1}{k-1}}$. Thus, if the adversary chooses a ``direct"
attack, by trying to recover the private exponent $m$, he will have
to solve the discrete log problem twice: first to recover
$\frac{k^{m}-1}{k-1}$ from $g^{\frac{k^{m}-1}{k-1}}$, and then to
recover $m$ from $k^{m}$. (Note that $k$ is public since $\phi$ is
public.)

On the other hand, the analog of what is called ``the Diffie-Hellman
problem" would be to recover the shared key $K =
g^{\frac{k^{m+n}-1}{k-1}}$ from the triple $(g,
~g^{\frac{k^{m}-1}{k-1}}, ~g^{\frac{k^{n}-1}{k-1}})$. Since $g$ and
$k$ are public, this is equivalent to recovering $g^{k^{m+n}}$ from
the triple $(g, ~g^{k^m},  ~g^{k^n})$, i.e., this is exactly the
standard Diffie-Hellman problem.

Thus, the bottom line of this example is that the instantiation of
our protocol where the group $G$ is $\mathbb{Z}_p^*$, is not really
different from the standard Diffie-Hellman protocol. In the next
section, we describe a more interesting instantiation, where the
(semi)group $G$ is non-commutative.

\section{Matrices over group rings and  extensions by inner automorphisms}
\label{Matrices}

To begin with, we note that our general protocol in Section
\ref{protocol_holomorph} can be used  with {\it any} non-commutative
group $G$  if $\phi$ is selected to be a non-trivial inner
automorphism, i.e., conjugation by an element which is not in the
center of $G$. Furthermore, it can be used  with  any
non-commutative {\it semigroup} $G$ as well, as long as  $G$ has
some invertible elements; these can be used to produce  inner
automorphisms. A typical example of such a semigroup would be a
semigroup of matrices over some ring.

In the paper \cite{KKS1}, the authors have employed matrices over
group rings of a (small) symmetric group as platforms for the
(standard) Diffie-Hellman-like key exchange. In this section, we use
these matrix semigroups again and consider an extension of such a
semigroup by an inner automorphism to get a platform semigroup for
our protocol.

Recall that a (semi)group ring $R[S]$ of a (semi)group  $S$ over a
commutative ring $R$ is the set of all formal sums

\[\sum_{g_i \in S} r_i g_i\]
where $r_i \in R$, and all but a finite number of $r_i$ are zero.

The sum of two elements in $R[G]$ is defined by

\[\left(\sum_{g_i\in S}a_ig_i\right)+\left(\sum_{g_i\in S}b_ig_i\right) = \sum_{g_i \in S}(a_i+b_i)g_i.\]

The multiplication of two elements in $R[G]$ is defined by using
distributivity.

As we have already pointed out, if a (semi)group $G$ is
non-commutative and has non-central invertible elements, then it
always has a non-identical inner automorphism, i.e., conjugation by
an element $g \in G$ such that $g^{-1} h g \ne h$ for at least some
$h \in G$.

Now let $G$ be the semigroup of  $3 \times 3$ matrices over the
group ring $\mathbb{Z}_{7}[A_5]$, where  $A_5$ is the alternating
group on 5 elements. 
Here we use an extension of the semigroup $G$ by an inner
automorphism $\varphi_{_H}$, which is conjugation by a matrix $H \in
GL_3(\mathbb{Z}_{7}[A_5])$. Thus, for any matrix $M \in G$ and for
any integer $k \ge 1$, we have

$$\varphi_{_H}(M) = H^{-1} M H; ~\varphi^k_{_H}(M) = H^{-k} M H^k.$$

\noindent Now our general protocol from Section
\ref{protocol_holomorph} is specialized in this case as follows.

\medskip

\begin{enumerate}

\item Alice and Bob agree on  public matrices $M \in G$ and $H \in
GL_3(\mathbb{Z}_{7}[A_5])$. Alice selects a private positive integer
$m$, and Bob selects a private positive integer $n$.
\medskip

\item Alice computes $(M, \varphi_{_H})^m = (H^{-m+1} M H^{m-1} \cdots H^{-2} M H^2 \cdot H^{-1} M H \cdot M,
~\varphi_{_H}^m)$ and sends {\bf only the first component} of this
pair
to Bob. Thus, she sends to Bob {\bf only} the matrix\\
$$A = H^{-m+1} M H^{m-1} \cdots H^{-2} M H^2 \cdot H^{-1} M H \cdot M
= H^{-m} (HM)^{m}.$$
\medskip

\item Bob computes $(M, \varphi_{_H})^n = (H^{-n+1} M H^{n-1} \cdots H^{-2} M H^2 \cdot H^{-1} M H \cdot M,
~\varphi_{_H}^n)$ and sends {\bf only the first component} of this
pair
to Alice. Thus, he sends to Alice {\bf only} the matrix\\
$$B = H^{-n+1} M H^{n-1} \cdots H^{-2} M H^2 \cdot H^{-1} M H \cdot M
= H^{-n} (HM)^{n}.$$


\item Alice computes $(B, x) \cdot (A, ~\varphi_{_H}^m) = (\varphi_{_H}^m(B) \cdot A,
~x  \cdot \varphi_{_H}^{m})$. Her key is now $K_{Alice} =
\varphi_{_H}^m(B) \cdot A = H^{-(m+n)}(HM)^{m+n}$. Note that she
does not actually ``compute" $x \cdot \varphi_{_H}^{m}$ because she
does not know the automorphism $x=\varphi_{_H}^{n}$; recall that it
was not transmitted to her. But she does not need it to compute
$K_{Alice}$.
\medskip

\item Bob computes $(A, y) \cdot (B, ~\varphi_{_H}^n) = (\varphi_{_H}^n(A) \cdot B,
~y \cdot \varphi_{_H}^{n})$. His key is now $K_{Bob} =
\varphi_{_H}^n(A) \cdot B$. Again, Bob does not actually ``compute"
$y \cdot \varphi_{_H}^{n}$ because he does not know the automorphism
$y=\varphi_{_H}^{m}$.
\medskip

\item Since $(B, x) \cdot (A, ~\varphi_{_H}^m) = (A, ~y) \cdot (B, ~\varphi_{_H}^n) =
(M, ~\varphi_{_H})^{m+n}$, we should have $K_{Alice} = K_{Bob} = K$,
the shared secret key.

\end{enumerate}

\section{Security assumptions and analysis}
\label{Security}

In this section, we address the question of security of the
particular instantiation of our protocol described in Section
\ref{Matrices}.

Recall that the shared secret key in the protocol of Section
\ref{Matrices} is

$$K = \varphi_{_H}^m(B) \cdot A = \varphi_{_H}^n(A) \cdot B = H^{-(m+n)}(HM)^{m+n}.$$

\noindent Therefore, our security assumption here is that it is
computationally hard to retrieve the key $K = H^{-(m+n)}(HM)^{m+n}$
from the quadruple\\
$(H, ~M, ~H^{-m}(HM)^{m}, ~H^{-n}(HM)^{n})$.

In particular, we have to take care that the matrices $H$ and  $HM$
do not commute because otherwise, $K$ is just a product of
$H^{-m}(HM)^{m}$  and  $H^{-n}(HM)^{n}$.

A weaker security assumption arises if an eavesdropper tries to
recover a private exponent from a transmission, i.e., to recover,
say, $m$ from $H^{-m} (HM)^{m}.$ A special case of this problem,
where $H=I$, is the ``discrete log" problem for matrices over
$\mathbb{Z}_{7}[A_5]$, namely: recover $m$ from $M$ and $M^{m}$.
Even this problem appears to be hard; it was addressed in
\cite{KKS1} in more detail. In particular, statistical experiments
show that for a  random matrix $M$, matrices $M^{m}$ are
indistinguishable from random.

In order to verify the robustness and security of our protocol, we
have experimentally  addressed two questions.  The first question is
whether or not any information about the private exponent $n$ is
leaked  from transmission. That is, for a random exponent $n$, how
different is the  matrix  $(M,\varphi_{_H})^n$ from $N$, where N is
random? The second point that needs verification is to determine how
different the final shared key is from a random matrix. More
specifically, if Alice and Bob choose secret integers $m$ and $n$
respectively, how different is the  matrix $(M, \varphi_{_H})^{n+m}$
from  $(M, \varphi_{_H})^q$, where $q$ is of the same bit size are
$n+m$.

To perform the first experimental validation we worked over
$M_3(\mathbb{Z}_7[A_5])$ and  used random choices of  $n \in
[10^{44},10^{55}]$. We then looked at the two distributions
generated by the first component of $(M, \varphi_{_H})^n$ and $N$,
where $M$ and  $N$ are  random matrices. We need to verify that the
two generated  distributions are in fact indistinguishable. To this
end we looked at the components of each matrix and counted the
frequency of occurrence of each element of $A_5$. We repeated this
process $500$ times and generated a frequency distribution table for
the two distributions.

From the table, we produced $Q-Q$ (quantile) plots of the entries of
the two matrices:  the first component of $(M, \varphi_{_H})^n$ and
a random matrix $N$. Quantile plots are a quick graphical tool for
comparing two distributions. These plots essentially compare the
cumulative distribution functions of two distributions. If the
distributions are identical, the resulting graph will be a straight
line.

\begin{figure}[!ht]
\includegraphics[width=0.95\textwidth]{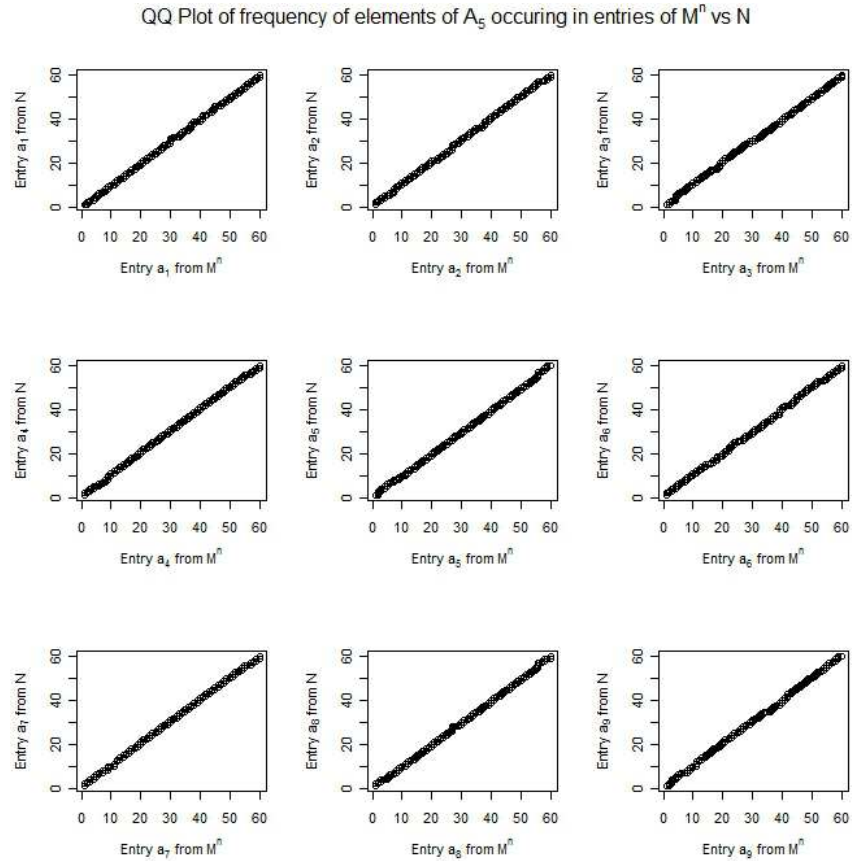}
\caption{Results for $M^{n}$ vs. $N$}
\label{semidirect_1}
\end{figure}

Figure \ref{semidirect_1} shows the resulting plots for this
experiment. These graphs show that the two distributions are in fact
identical, therefore suggesting that no information about a private
exponent $n$ is revealed by transmissions between Alice and Bob.

\begin{figure}[!ht]
\includegraphics[width=0.95\textwidth]{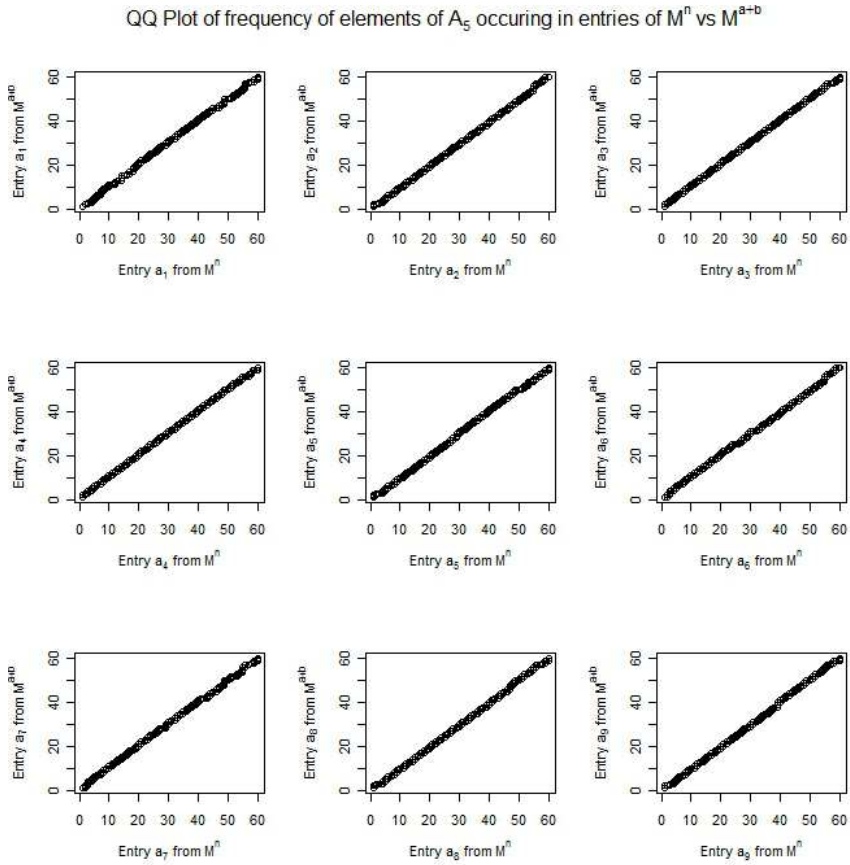}
\caption{Results for $M^{n}$ vs. $M^{a+b}$} \label{semidirect_2}
\end{figure}

The second experiment we carried out was similar to the first one,
except in this case  we were comparing the first components of $(M,
\varphi_{_H})^n$ and $(M, \varphi_{_H})^{a+b}$, where $n, a$ and $b$
are random and all of roughly the same bit size, i.e. all are
integers from $[10^{44},10^{55}]$. This experiment helps address the
DDH (decisional Diffie-Hellman) assumption by comparing the shared
secret key to a random key and ensuring that no information about
the former is leaked. See Figure \ref{semidirect_2} for the
resulting $Q-Q$ plots. These 9 graphs suggest that the two
distributions generated by these keys are in fact indistinguishable.

\section{Parameters and key generation}
\label{Parameters}

Private exponents $m$ and $n$ should be of the magnitude of $2^t$,
where $t$ is the security parameter, to make brute force search
infeasible. Thus, $m$ and $n$ are roughly $t$ bits long.

Public matrix $M$ is selected as a random $3 \times 3$ matrix over
the group ring $\mathbb{Z}_{7}[A_5]$, which means that each  entry
of $M$ is a random element of $\mathbb{Z}_{7}[A_5]$. The latter
means that each  entry is a sum $\sum_{g_i \in A_5} c_i g_i$ of
elements of the group  $A_5$ with coefficients $c_i$ selected
uniformly randomly from $\mathbb{Z}_{7}$. Thus, although the bit
complexity of the matrix $M$ is fairly high ($9 \cdot 3 \cdot 60 =
1620$ bits), the procedure for sampling $M$ is quite efficient. We
want to impose one restriction on the matrix $M$ though. There is a
trivialization (sometimes called {\it augmentation}) homomorphism of
the group ring that sends every group element to 1. This
homomorphism naturally extends to a homomorphism of the whole
semigroup of matrices. To avoid leaking any information upon
applying this homomorphism, we want the image of every entry of $M$
to be 0. Group ring elements like that are easy to sample: after
sampling a random element $\sum_{g_i \in A_5} c_i g_i$ of
$\mathbb{Z}_{7}[A_5]$, we select a random coefficient $c_i$ and
change it, if necessary, to have $\sum_i c_i = 0.$

Note that with this choice of $M$, applying the trivialization
homomorphism to any of the transmitted matrices in our protocol will
produce the zero matrix, thus not leaking any information. We also
note that there are no other homomorphisms of the group $A_5$ (which
is a finite simple group), except for inner automorphisms. This will
prevent an eavesdropper from learning partial information about
secret keys by applying homomorphisms to transmitted matrices.

Finally, we need to sample an {\it invertible} $3 \times 3$ matrix
$H$ over the group ring $\mathbb{Z}_{7}[A_5]$. There are several
techniques for doing this; here we give a brief exposition of one
possible procedure.

We start with an already ``somewhat random" matrix, for which it is
easy to compute the inverse. An example of such a matrix is a
lower/upper triangular matrix, with invertible elements on the
diagonal:
\begin{align*}
U&=\begin{pmatrix}
g_1 & u_1 & u_2 \\
0 & g_2 & u_3 \\
0 & 0 & g_3 \\
\end{pmatrix}.
\end{align*}

Here $g_i$ are random elements of the group $A_5$, and $u_i$ are
random elements of the group ring $\mathbb{Z}_{7}[A_5]$. We then
take a random product, with 20 factors, of such random  invertible
upper and lower triangular matrices, to get our invertible  matrix
$H$.

We note that there is always a concern (also in the standard
Diffie-Hellman protocol) about the order of a public element: if the
order is too small, then a brute force attack may be feasible. In
our situation, this concern is significantly alleviated by the fact
that our transmissions are products of powers of two different
matrices rather  than  powers of a single matrix. Therefore, even if
the order of one of the matrices happens to be small by accident,
this does not mean that the product $H^{-m} (HM)^{m}$ will go into
loop of a small size. Furthermore, since our matrix $M$ is
non-invertible,   it does not have an ``order", but rather a loop:
$M^r=M^s$ for some positive  $r \ne s$. The matrices $HM$ and
$H^{-m} (HM)^{m}$ are non-invertible, too, so they do not have an
order either, but rather a loop. Detecting a loop is, in general,
computationally much harder than computing the order of an
invertible element.

\section{Conclusions}

We have presented  a brand new  key exchange protocol based  on
extension of a (semi)group by automorphisms and  described some
practical instances of this general idea. Our protocol can be based
on any group, in particular on any non-commutative group. It has
some superficial resemblance to the classical Diffie-Hellman
protocol, but there are several distinctive features that, we
believe, give our protocol important advantages:

$\bullet$ Even though the parties do compute a large power of a
public element (as in the classical Diffie-Hellman protocol), they
do not transmit the whole result, but rather just part of it.

$\bullet$ Since the classical Diffie-Hellman protocol is a special
case of our protocol, breaking our protocol even for any cyclic
group would imply breaking the  Diffie-Hellman protocol.

$\bullet$ If the platform (semi)group is not commutative, then we
get a new security assumption. In the simplest case, where the
automorphism used for extension is inner, attacking a private
exponent amounts to recovering an integer $n$ from a product
$g^{-n}h^{n}$, where  $g, h$ are public elements of the platform
(semi)group. In the special case where $g=1$ this boils down to
recovering $n$ from $h^{n}$, with public $h$ (``discrete log"
problem).

On the other hand, in the particular instantiation of our protocol,
which is based on a non-commutative semigroup extended by an inner
automorphism, recovering the shared secret key from public
information is based on a different security assumption than the
classical Diffie-Hellman protocol is. Namely, the assumption is that
it is computationally hard to retrieve the shared secret key $K =
h^{-(m+n)}g^{m+n}$ from the triple of elements $(h, ~h^{-m} g^{m},
~h^{-n}g^{n})$, assuming that  $g$ and $h$ do not commute.


%

\end{document}